\documentclass[12pt,english]{article}

\usepackage{geometry}
\usepackage{amsthm, amssymb}
\usepackage[titletoc,title]{appendix}

\usepackage{setspace}
\def\ket#1{\left|#1\right>}
\def\bra#1{\langle#1\vert}
\def\exptt#1{\langle#1\rangle}
\def\expt(#1#2){\langle #2 \vert #1 \vert #2 \rangle}

\newcommand{\I}{{{i}\over{\hbar}}}
\newcommand{\E}{\text{e}}

\usepackage{sectsty}
\allsectionsfont{\normalsize\raggedright\centering}

\usepackage[]{tocloft}
\addtocontents{toc}{\cftpagenumbersoff{section}}
\addtocontents{toc}{\cftpagenumbersoff{subsection}}

\makeatletter
\renewcommand{\@cftmaketoctitle}{}
\makeatother

\makeatletter
\renewcommand\@biblabel[1]{}
\makeatother

\usepackage[]{footmisc}

\usepackage{url}
\makeatletter
\def\url@leostyle{%
    \def\UrlFont{\sf}}{\def\UrlFont{\small\ttfamily}}
\makeatother
\usepackage{graphicx}
\usepackage{babel}

\usepackage{amsmath}
\usepackage{amsfonts}

\usepackage{framed}

\usepackage{xifthen}


\begin{document}

\title{{\LARGE \bf{Protective Measurements and the Reality of the Wave Function}}}
\author{{\large \bf{Shan Gao}}}
\date{}

\maketitle

\thispagestyle{empty}

\begin{abstract}
It has been debated whether protective measurement implies the reality of the wave function. In this paper, I present a new analysis of the relationship between protective measurements and the reality of the wave function. First, I briefly introduce protective measurements and the ontological models framework for them. Second, I give a simple proof of Hardy's theorem in terms of protective measurements. Third, I analyze two suggested $\psi$-epistemic models of a protective measurement. It is shown that although these models can explain the appearance of expectation values of observables in a single measurement, their predictions about the variance of the result of a non-ideal protective measurement are different from those of quantum mechanics. Finally, I argue that under an auxiliary finiteness assumption about the dynamics of the ontic state, protective measurement implies the reality of the wave function in the ontological models framework. 

\end{abstract}

\mbox{} \\

\tableofcontents

\mbox{} \\

\section{Introduction}

The reality of the wave function has been a hot topic of debate since the early days of quantum mechanics. Recent years have witnessed a growing interest in this long-standing question. 
Is the wave function real, directly representing the ontic state of a physical system, or epistemic, merely representing a state of incomplete knowledge about the underlying ontic state? 
A general and rigorous approach called ontological models framework has been proposed to distinguish the $\psi$-ontic and $\psi$-epistemic views and determine the relation between the wave function and the ontic state of a physical system, such as whether a given ontic state is compatible with two distinct wave functions (Spekkens [2005]; Harrigan and Spekkens [2010]). 
Moreover, several $\psi$-ontology theorems have been proved in the framework (Pusey, Barrett and Rudolph [2012]; Colbeck and Renner [2012], [2017]; Hardy [2013]). In particular, the Pusey-Barrett-Rudolph theorem shows that in the ontological models framework, when assuming independently prepared systems have independent ontic states, the ontic state of a physical system uniquely determines its wave function, and thus the wave function directly represents the ontic state of the system (Pusey, Barrett and Rudolph [2012]).  

However, a definite answer to the above question is still unavailable. 
On the one hand, auxiliary assumptions are required to prove the existing $\psi$-ontology theorems, such as the preparation independence assumption for the Pusey-Barrett-Rudolph theorem (Pusey, Barrett and Rudolph [2012]), the freedom of choice assumption for the Colbeck-Renner theorem (Colbeck and Renner [2012], [2017]), and the ontic indifference assumption for Hardy's theorem (Hardy [2013]). 
It thus seems impossible to completely rule out the $\psi$-epistemic view without auxiliary assumptions. Indeed, by removing these auxiliary assumptions, explicit $\psi$-epistemic models can be constructed to reproduce the statistics of quantum mechanics for projective measurements in orthonormal bases in Hilbert spaces of any dimension (Lewis et al [2012]; Aaronson et al [2013]). 
However, these models do not reproduce the quantum predictions for all possible measurements such as POVMs. 
As Leifer ([2014]) rightly pointed out, ``it is still possible that there are no $\psi$-epistemic models that reproduce the quantum predictions for all POVMs, and it may be possible to prove this without auxiliary assumptions.''

On the other hand, it has been known that there are other types of quantum measurements besides the conventional projective measurements, such as weak measurements and protective measurements (Aharonov and Vaidman [1993]; Aharonov, Anandan and Vaidman [1993]; Piacentini et al [2017]). 
Moreover, it has been conjectured that protective measurements, which can measure the expectation values of observables and even the wave function on a single quantum system, provide strong supports for the reality of the wave function (Aharonov and Vaidman [1993]; Aharonov, Anandan and Vaidman [1993], [1996]; Gao [2014], [2015] [2017]; Hetzroni and Rohrlich [2014]). 
However, it has also been argued that this is not the case (Unruh [1994]; Rovelli [1994]; Dass and Qureshi [1999]; Schlosshauer and Claringbold [2014]; Combes et al [2018]). Thus it is still controversial whether protective measurement really implies the reality of the wave function. 

In this paper, I will present a new analysis of the relationship between protective measurements and the reality of the wave function. In particular, I will give a new proof of the reality of the wave function in terms of protective measurements in the  ontological models framework. Like the existing $\psi$-ontology theorems such as the Pusey-Barrett-Rudolph theorem, the proof also relies on an auxiliary assumption. 

The rest of this paper is organized as follows. 
In Section 2, I first give a brief introduction to protective measurements (PMs).  
It is emphasized that PM is a natural result of the Schr\"{o}dinger equation; when the wave function of the measured system is protected to be unchanged during a standard von Neumann measurement of an observable, the result is naturally the expectation value of the observable in the wave function of the measured system.  
Besides, I also briefly introduce two known schemes of PM: the adiabatic-type PM or A-PM and the Zeno-type PM or Z-PM.  
In Section 3, I then introduce the ontological models framework, which provides a  general and rigorous approach to determine whether the wave function is ontic or epistemic. In particular, I introduce an important assumption of the framework for PMs, namely a rule of connecting the underlying ontic states with the results of PMs, which says that the definite result of a PM is determined by the total evolution of the ontic state of the protected system during the PM. 
In Section 4, I derive a basic result based on the ontological models framework for PMs, which is that two protected wave functions correspond to different evolution of the ontic state of the protected system during a PM. 

In Section 5, I take Hardy's theorem as an example to show that PM may have implications for the reality of the wave function in the ontological models framework. 
The key assumption of Hardy's theorem is the ontic indifference assumption, which says that any quantum transformation on a system which leaves its wave function unchanged (including those of PMs) can be performed in such a way that it does not affect the ontic state of the system. I argue that PM provides a simple proof of Hardy's theorem under the ontic indifference assumption. 
In Section 6, I turn to the dynamics of the ontic state during a PM by analyzing two suggested $\psi$-epistemic models of a PM (one for a Z-PM and the other for an A-PM), in which the ontic state of the system is affected by the PM. 
It is shown that although these models can explain the appearance of expectation values of observables in a single measurement, their predictions about the variance of the result of a non-ideal PM are different from those of quantum mechanics.\footnote{In this paper, when I say a PM I usually mean an ideal PM which yields a definite result unless stated otherwise. Sometimes I also say ideal PM, and this is emphasis.} 
In Section 7, I argue that under a weaker and more natural finiteness assumption about the dynamics of the ontic state, PM implies the reality of the wave function in the ontological models framework.  
Conclusions are given in the last section. 




\section{Protective measurements}

Protective measurement (PM) is a method to measure the expectation value of an observable on a single quantum system (Aharonov and Vaidman [1993]; Aharonov, Anandan and Vaidman [1993]; Vaidman [2009]; Gao [2014]). 
For a conventional projective measurement, the wave function of the measured system is in general changed greatly during the measurement, and one obtains an eigenvalue of the measured observable randomly, and the expectation value of the observable can be obtained only as the statistical average of eigenvalues for an ensemble of identically prepared systems. 
By contrast, during a PM the wave function of  the measured system is protected by an appropriate procedure so that it keeps unchanged during the measurement. 
Then, by the Schr\"{o}dinger evolution, the measurement result will be directly the expectation value of the measured observable, even if the system is initially not in an eigenstate of the observable. 

This result can be seen clearly from the following simple derivation. 
As for a projective measurement, the interaction Hamiltonian for measuring an observable $A$ is given by the usual form $H_I = g(t)PA$, where $g(t)$ is the time-dependent coupling strength of the interaction, which is a smooth function normalized to $\int_0^T g(t)dt =1$ during the measurement interval $T$, and $g(0)=g(T)=0$, and $P$ is the conjugate momentum of the pointer variable $X$. 
When the wave function of the measured system is protected to keep unchanged during the measurement, the evolution of the wave function of the combined system is

\begin{equation}
\ket{\psi(0)}\ket{\phi(0)} \rightarrow \ket{\psi(t)}\ket{\phi(t)}, t>0,
\end{equation}

\noindent where $\ket{\phi(0)}$ and $\ket{\phi(t)}$ are the wave functions of the measuring device at instants $0$ and $t$, respectively, $\ket{\psi(0)}$ and $\ket{\psi(t)}$ are the wave functions of the measured system at instants $0$ and $t$, respectively, and $\ket{\psi(t)}$ is the same as $\ket{\psi(0)}$ up to an overall phase during the measurement interval $[0,T]$.\footnote{Note that since this phase can also be attributed to the state of the measuring device, we can keep $\ket{\psi(t)}=\ket{\psi(0)}$.} Then we have

\begin{eqnarray}
{d \over dt}\bra{\psi(t)\phi(t)}X\ket{\psi(t)\phi(t)} &=& {1\over {i\hbar}}\bra{\psi(t)\phi(t)}[X, H_I]\ket{\psi(t)\phi(t)}\nonumber\\
&=&g(t)\bra{\psi(0)}A\ket{\psi(0)}.
\end{eqnarray}

\noindent Note that the momentum expectation value of the pointer is zero at the initial instant and the free evolution of the pointer conserves it. This further leads to

\begin{equation}
\bra{\phi(T)}X\ket{\phi(T)}- \bra{\phi(0)}X\ket{\phi(0)}= \bra{\psi(0)}A\ket{\psi(0)},
\end{equation}

\noindent which means that the shift of the center of the pointer wavepacket is the expectation value of $A$ in the initial wave function of the measured system.  
This clearly demonstrates that the result of a measurement of an observable on a  system, which does not change the wave function of the system, is the expectation value of the measured observable in the wave function of the measured system. 


There are two known schemes of PM (Aharonov and Vaidman [1993]; Aharonov, Anandan and Vaidman [1993]). 
The first scheme is to introduce a protective potential such that the wave function of the measured system at a given instant, $\ket{\psi}$, is a nondegenerate energy eigenstate of the total Hamiltonian of the system with finite gap to neighboring energy eigenstates. By this scheme, the measurement of an observable is required to be weak and adiabatic. We may call this scheme adiabatic-type PM or A-PM. 
An ideal A-PM requires $T\Delta E \rightarrow \infty$, where $T$ is the measurement interval, and $\Delta E$ is the smallest of the energy differences between $\ket{\psi}$ and other energy eigenstates.\footnote{There are two types of ideal A-PMs. The first is the usual type which makes $T \rightarrow \infty$. The second is to make $\Delta E \rightarrow \infty$ and $T$ large enough but finite (so that the interaction Hamiltonian can be regarded as a perturbation). The second type of ideal A-PMs is like ideal Z-PMs, and $\Delta E$ plays the similar role of $N$ in Z-PMs.} 
The second scheme is via the quantum Zeno effect, and it may be called Zeno-type PM or Z-PM. 
The Zeno effect is realized by making frequent projective measurements of an observable, of which the wave function of the measured system at a given instant, $\ket{\psi}$, is a nondegenerate eigenstate. 
By this scheme, the measurement of the measured observable is not necessarily weak but weaker than the Zeno projective measurements. 
An ideal Z-PM requires $N \rightarrow \infty$, where $N$ is the number of Zeno projective measurements. 

Since the wave function can be reconstructed from the expectation values of a sufficient number of observables, the wave function of a single quantum system can be measured by a series of PMs (which are performed in parallel at the same time).
Then, it seems natural to conjecture that the wave function refers directly to the physical state of the system. 
As noted before, however, there have been concerns about the validity of the conjecture. 
For one, PM cannot measure an arbitrary unknown wave function. Rather, it requires  some information about the measured wave function in order to provide the protection.   
This permits the possibility that what a PM measures may be the protection procedure, not the system itself. 
On the other hand, it has been argued that since we can use many (or even infinitely many) different protection procedures to obtain the same wave function, what a PM measures is not the protection procedure, but the system itself (Aharonov, Anandan and Vaidman [1996]). 
No doubt, in order to investigate whether the above conjecture is true, heuristic arguments are not enough, and we need a more rigorous approach. 

\section{Ontological models framework}

A general and rigorous approach to determine whether the wave function is ontic or epistemic is the ontological models framework (Spekkens [2005]; Harrigan and Spekkens [2010]; Leifer [2014]). It has two fundamental assumptions. 

The first assumption is about the existence of the underlying state of reality. It says that if a physical system is prepared such that quantum mechanics assigns a wave function to it, then after preparation the system has a well-defined set of physical properties or an underlying ontic state, which is usually represented by a mathematical object, $\lambda$. 
In general, for an ensemble of identically prepared systems to which the same wave function $\psi$ is assigned, the ontic states of different systems in the ensemble may be different, and the wave function $\psi$ corresponds to a probability distribution $p(\lambda|\psi)$ over all possible ontic states, where $\int{d\lambda p(\lambda|\psi)}=1$. 
Here a strict $\psi$-ontic/epistemic distinction can be made. 
In a $\psi$-ontic model, the ontic state of a physical system uniquely determines its wave function, and the probability distributions corresponding to two different wave functions do not overlap. 
In this case, the wave function directly represents the ontic state or a property of the system.\footnote{In a $\psi$-ontic model, the wave function is not necessarily complete; that is to say, it does not necessarily represent the complete ontic state of a system, such as in Bohm's theory.} 
While in a $\psi$-epistemic model, there are at least two wave functions which are compatible with the same ontic state of a physical system, and the probability distributions corresponding to two different wave functions may overlap. 
In this case, the wave function represents a state of incomplete knowledge - an epistemic state - about the actual ontic state of the system. 


In order to investigate whether an ontological model is consistent with the empirical predictions of quantum mechanics, we also need a rule of connecting the underlying ontic states with the results of measurements. 
This is the second assumption of the ontological models framework, which says that when a measurement is performed, the behaviour of the measuring device is determined only by the ontic state of the system, along with the physical properties of the measuring device. 
For a projective measurement $M$, its result is random in general. 
Then this assumption means that the ontic state $\lambda$ of a physical system determines the probability $p(k|\lambda,M)$ of different results $k$ for the measurement $M$ on the system. 
The consistency with the predictions of quantum mechanics requires the following relation: $\int{d\lambda p(k|\lambda, M)p(\lambda|\psi)} = p(k|M, \psi)$, where $p(k|M, \psi)$ is the Born probability of $k$ given $M$ and $\psi$. 

For a PM, which yields a definite measurement result, it seems that the above assumption should mean that the ontic state of a physical system determines the definite result of the PM on the system (Gao [2015]). 
The behaviour of the measuring device during a PM is yielding the unique, definite result of the PM after all. 
However, different from a projective measurement, the ontic state of the measured system may be affected by the protection procedure during a PM (Combes et al [2018]), and thus it seems not reasonable to assume that when a PM is performed, the behaviour of the measuring device is determined by the ontic state of the measured system (along with the physical properties of the measuring device) immediately before the PM. 
A more reasonable assumption for PMs is that the ontic state of the measured system may be affected by the protection procedure and thus evolve in a certain way during a PM, and the definite result of the PM is determined by the total evolution of the ontic state of the system during the PM, not simply by the initial ontic state of the system (see also Gao [2017]). 

This assumption is in accordance with the consistency condition for PMs.  
For a PM of an observable $A$, the consistency with the predictions of quantum mechanics requires the relation $\int{d\lambda(t) p(k|\lambda(t), M)p(\lambda(t)|P)} = p(k|M, P)$, where $P$ is the preparation of the protected system by a protection procedure, $M$ is a usual measurement of $A$, $\lambda(t)$ denotes the total evolution of the ontic state of the protected system during the PM, $p(k|M, P)=\delta(k-\exptt{A})$, $\exptt{A}$ is the unique, definite result of the PM. 
It can be seen that this consistency condition requires that $p(k|\lambda(t), M)=\delta(k-\exptt{A})$ for all possible $\lambda(t)$, which means that the total evolution of the ontic state of the protected system during the PM determines the definite result of the PM. 

Here it may be worth noting that the above ontological models framework also has limitations, and in particular, the assumptions of the framework are not universially accepted by all quantum theories. For example, a QBist or Healey-style pragmatist may insist that the wave function is epistemic, while denying that there is an underlying ontic state (Fuchs, Mermin and Schack [2014]; Healey [2017]).

\section{A basic result}

In the following sections, I will analyze whether PM has implications for the reality of the wave function in the above ontological models framework. I will first derive a basic result and then make it stronger by resorting to an auxiliary assumption. 

For any two protected states $\ket{\psi_1}$ and $\ket{\psi_2}$, which are prepared by the protection procedures $P_1$ and $P_2$, respectively, we can choose an observable $A$ whose expectation values in the two states are different. 
Consider a measurement $M$ of this observable $A$ on each of these two protected states. 
Suppose the two states $\ket{\psi_1}$ and $\ket{\psi_2}$ correspond to two probability distributions $p(\lambda_1(t)|P_1)$ and $p(\lambda_2(t)|P_2)$, where $\lambda_1(t)$ and  $\lambda_2(t)$ are possible evolution of the ontic state of the protected system during the measurement interval when the protection procedures are $P_1$ and $P_2$.  
According to the above analysis, for the same measurement $M$ of $A$ on the two protected states, we have $p(k|\lambda_1(t), M)=\delta(k-\exptt{A}_{\psi_1})$ and $p(k|\lambda_2(t), M)=\delta(k-\exptt{A}_{\psi_2})$, where $\exptt{A}_{\psi_1}$ and $\exptt{A}_{\psi_2}$ are the expectation values of $A$ in the two protected states $\ket{\psi_1}$ and $\ket{\psi_2}$, respectively. 
Since $\exptt{A}_{\psi_1} \neq \exptt{A}_{\psi_2}$, we find that $\lambda_1(t) \neq \lambda_2(t)$ and thus the two probability distributions $p(\lambda_1(t)|P_1)$ and $p(\lambda_2(t)|P_2)$ do not overlap with one another.   
In other words, two protected wave functions correspond to different evolution of the ontic state of the protected system during the measurement interval of a PM. 

This result can also be proven by reduction to absurdity. 
For two protected wave functions, choose an observable whose expectation values in these two states are different. 
Then the same measurement of this observable on these two protected wave functions will yield two different results with certainty. 
If there exists a probability $p>0$ that these two wave functions correspond to the same evolution of the ontic state of the protected system, $\lambda(t)$, during the measurement interval, then since $\lambda(t)$ determines the definite result of each measurement according to the ontological models framework, the results of the measurements of the observable on these two states will be the same with probability not smaller than $p$. 
This leads to a contradiction. 
Therefore, two protected wave functions of a system correspond to different evolution of the ontic state of the system during the measurement interval of a PM. 
This also means that the total evolution of the ontic state of a protected system during a PM uniquely determines the wave function of the system. 

In fact, this result can be obtained more directly from the second assumption of the ontological models framework for PMs.  
According to this assumption, the total evolution of the ontic state of a protected system during a PM determines the definite result of the PM, namely the expectation value of the measured observable which may be arbitrary. 
Since a wave function can be constructed from the expectation values of a sufficient number of observables, the total evolution of the ontic state of a protected system during a PM also determines the wave function of the system. 

Recall that there are two possible types of models in the ontological models framework. 
In a $\psi$-ontic model, the wave function is determined by the ontic state, and it represents a property of a physical system. 
While in a $\psi$-epistemic model, the wave function is not determined by the ontic state, and it does not represent a property of a physical system. 
Similarly, when the wave function of a protected system is determined by the total evolution of the ontic state of the system during a PM, we may say that 
the wave function represents a property of the protected system during the measurement interval of the PM (see also Aharonov, Anandan and Vaidman [1996]). 
This is a basic result derived from the ontological models framework for PMs. 

As noted before, during a PM, the ontic state of the protected system may be disturbed by the protection procedure and thus its total evolution may be determined not only by the initial ontic state, but also by the protection procedure. 
Thus, the above result does not imply the reality of the wave function.  
In order to establish the reality of the wave function, we need to further prove that the wave function of a protected system is determined only by the initial ontic state of the system, or the result of a PM is determined only by the initial ontic state of the system. 
It seems that we must resort to auxiliary assumptions to prove this.  

\section{A simple proof of Hardy's theorem} 

In this section, I will use Hardy's theorem as an example to show that when resorting to auxiliary assumptions we can prove the reality of the wave function in the ontological models framework for PMs. 

Hardy's theorem is one of the three important $\psi$-ontology theorems appeared in recent years (Hardy [2013]).  
It is based on three assumptions. The first one is realism, which says that each time a system is prepared there exists an underlying state of reality or an ontic state, denoted by $\lambda$. This is just the first assumption of the ontological models framework. The second assumption of Hardy's theorem is possibilistic completeness, which says that the ontic state, $\lambda$, is sufficient to determine whether any outcome of any (projective) measurement has probability equal to zero of occurring or not. This is a weaker version of the second assumption of the ontological models framework, according to which the ontic state determines the probabilities for the results of projective measurements.  
The third assumption of Hardy's theorem is an auxiliary assumption and also the key assumption of the theorem, called ontic indifference, which says that any quantum transformation on a system which leaves unchanged any given wave function $\ket{\psi}$ can be performed in such a way that no underlying ontic state which is assigned a nonzero probability by $\ket{\psi}$ is affected. Hardy's theorem then states that under the above three assumptions any pair of wave functions must have non-overlapping distributions over the ontic states and thus the wave function is real, directly representing the ontic state of a single quantum system (Hardy [2013]). 

Hardy's theorem can be illustrated with a simple example (Leifer [2014]). Assume two nonorthogonal states $\ket{\psi_1}$ and ${1 \over \sqrt{2}}(\ket{\psi_1}+\ket{\psi_2})$ are compatible with the same ontic state $\lambda$, where $\ket{\psi_1}$ is orthogonal to $\ket{\psi_2}$. Consider a unitary evolution which leaves $\ket{\psi_1}$ invariant but changes ${1 \over \sqrt{2}}(\ket{\psi_1}+\ket{\psi_2})$ to its orthogonal state ${1 \over \sqrt{2}}(\ket{\psi_1}-\ket{\psi_2})$. Since two orthogonal states correspond to different ontic states,\footnote{Note that the possibilistic completeness assumption is needed to prove this result.} the original ontic state $\lambda$ must be changed by the unitary evolution. Then if the unitary evolution that leaves $\ket{\psi_1}$ invariant also leaves the underlying ontic state $\lambda$ invariant as the ontic indifference assumption requires,\footnote{One strong motivation for this assumption  is locality. When $\ket{\psi_1}$ and $\ket{\psi_2}$ are two spatially separated states prepared in regions 1 and 2 respectively, it seems reasonable to assume that the local evolution of the ontic state in region 2 does not influence the ontic state in region 1.}  there will be a contradiction. In other words, under the above three assumptions we can prove that the two nonorthogonal state $\ket{\psi_1}$ and ${1 \over \sqrt{2}}(\ket{\psi_1}+\ket{\psi_2})$ are ontologically distinct. 

This is the simplest example of Hardy's theorem. A complete proof of this theorem requires a more complex mathematical analysis. 
In the following, I will show that under the key assumption of Hardy's theorem, namely the ontic indifference assumption, PM implies the reality of the wave function in the ontological models framework. This will provide a simple proof of Hardy's theorem. 

Before giving the proof, I should first point out that the ontic indifference assumption in Hardy's theorem is a very strong assumption on a $\psi$-epistemic view (Leifer [2014]; Combes et al. [2018]). On this view, as I have introduced before, the wave function does not correspond to the ontic state, and thus it is possible that the underlying ontic state changes even if the wave function stays the same. In this sense, my following proof of Hardy's theorem should be regarded only as a suggestion that PM may have implications for the reality of the wave function under certain auxiliary assumptions.

Here is the proof. 
First, according to the basic result obtained above, 
the wave function of a protected system is determined by the total evolution of the ontic state of the system during a PM. 
Next, the ontic indifference assumption implies that the PM (which keeps the wave function of the system unchanged) can be performed in such a way that the ontic state of the system is not changed. 
This means that the ontic state of the system at each instant during the PM is the same as the initial ontic state. 
Then, the wave function of the system is determined by the initial ontic state of the system before the PM. 
In other words, the wave function of an (unprotected) system is determined by the ontic state of the system, and it represents the ontic state of the system according to the $\psi$-ontic/epistemic distinction. 
This establishes the reality of the wave function and proves Hardy's theorem. 

Finally, it is worth noting that Hardy's theorem can also be proven under the restricted ontic indifference assumption, namely the theorem can be proven even if the ontic indifference assumption holds only for a single wave function (Hardy [2013]; Patra, Pironio and Massar [2013]). 
However, the above proof in terms of PMs cannot go through if the ontic indifference assumption holds only for a single wave function; in this case, the proof will only  establish the reality of this wave function. 

\section{On two $\psi$-epistemic models of a PM}

The above analysis shows that when assuming the ontic state of a protected system keeps unchanged during a PM, the reality of the wave function can be proved. 
Then, a $\psi$-epistemic model must assume that the ontic state of a protected system evolves over time in a certain way in order to account for PMs. 
Concretely speaking, the ontic state of the protected system must undergo a dynamical process to generate the result of the PM, namely the expectation value of the measured observable. 
The question is: can any dynamics of the ontic state account for PMs? 
In this section, I will analyze two recently suggested $\psi$-epistemic models of PMs, one for Z-PMs and the other for A-PMs (Combes et al [2018]). 

For a Z-PM, there is an ensemble of identically prepared copies of the measured system, which is prepared by the protection procedure, namely the frequent Zeno projective measurements, when the protection is successful. 
Thus, it seems possible that the result of the Z-PM, namely the expectation value of the measured observable, is also obtained as the ensemble average of the eigenvalues of the measured observable as for conventional projective measurements. 
Indeed, Combes et al (2018) suggested such a $\psi$-epistemic model for a Z-PM.\footnote{The model discussed below is an extension of the original model for a spin-1/2 particle.} The model assumes that any observable $A$ of the measured system has a definite value at any time, which is one of the eigenvalues of $A$. Similarly, the pointer of the measuring device also has a definite position at any time, which is the same as the measured position predicted by quantum mechanics.  
Moreover, when each Zeno projective measurement results in the wave function of the measured system being in $\ket{\psi}$, it randomizes the value of $A$ and makes it be $a_i$ with probability $p_i$, where $a_i$ is an eigenvalue of $A$, and 
$p_i=|\langle{a_i}|{\psi}\rangle|^2$ is the corresponding Born probability. 
Then the measured system shifts the pointer by $a_i/N$ after the follow-up measurement of $A$. 
In the end, the total pointer shift, denoted by $\Delta x$, will be the expectation value of $A$ when $N$ approaches infinity:

\begin{equation}
\Delta x = \lim_{N \rightarrow \infty}\sum_i{n_ia_i/N} = \sum_i{p_ia_i} = \exptt{A}.
\label{xxx}
\end{equation}

This $\psi$-epistemic model shows that the result of a Z-PM, the expectation value of the measured observable, may be generated from the eigenvalues of the observable for an ensemble of identically prepared copies of the measured system, which is prepared by the protection procedure in the Z-PM. 

However, as Combes et al (2018) also pointed out, the model does not aim to provide a complete account of a Z-PM, which means that the predictions of the model may be not fully consistent with those of quantum mechanics.  
This is indeed the case, since it can be shown that this $\psi$-epistemic model and quantum mechanics give different predictions about the variance of the result of a Z-PM with finite $N$. 

A Z-PM with finite $N$ is composed of $N$ identical units, each of which contains a protecting system and a measuring system. 
In the above $\psi$-epistemic model, the pointer shift generated by the $i$-th Z-PM unit, $\Delta x_i$, has a probability distribution 

\begin{equation}
p(\Delta x_i=a_k) = |\langle{a_k}|{\psi}\rangle|^2.
\label{}
\end{equation}

\noindent Thus we have $Var(\Delta x_i)=Var(A)/N^2$ for any $i$, where $Var(\cdot)$ is the variance, and $Var(A) \equiv \exptt{A^2}-\exptt{A}^2 $. Then the variance of the final position of the pointer after the Z-PM is

\begin{equation}
Var(x_f) = Var(x_0+\sum_i{\Delta x_i}), 
\label{}
\end{equation}

\noindent where $x_f$ is the final position of the pointer, and $x_0$ is the initial  position of the pointer. 
Since each random process $\Delta x_i$ is independent with each other and also independent of the initial position of the pointer in the model, we have

\begin{equation}
Var(x_f) = Var(x_0)+Var(\sum_i{\Delta x_i})=Var(x_0)+{{Var(A)}\over{N}}.  
\label{}
\end{equation}

On the other hand, according to quantum mechanics, the branch of the state of the combined system after the Z-PM (namely after $N$ such measurements), in which each Zeno projective measurement results in the state of the measured system being in $\ket{\psi}$, is (up to the first order of 1/N)

\begin{eqnarray}
  \ket{t=T} 
&=& \ket{\psi}\ket{\phi(x_0+ \exptt{A})} \nonumber \\ & & + {{Var(A)}\over {2N}}\ket{\psi} \ket{\phi''(x_0+\exptt{A})},
\label{pm}
\end{eqnarray}

\noindent  where  $\phi(x_0)$  is the initial pointer wavepacket. Suppose the initial pointer wavepacket is a Gaussian wavepacket. Then we can calculate the variance of the final measuerd position of the pointer, which is
 
 \begin{equation}
Var(x_f)=Var(x_0) +{{Var(A)}\over{N}}Var(x_0)(k_1+k_2Var(x_0)),
\label{xfqm}
\end{equation}

\noindent where $Var(x_0)$ is the variance of the initial measured position of the pointer, and $k_1$, $k_2$ are numerical constants related to the Gaussian wavepacket. 

It can be seen that the above $\psi$-epistemic model and quantum mechanics give obviously different predictions about the variance of the result of a Z-PM with finite $N$. In the model, the first order term does not depend on the initial position variance of the pointer, but in quantum mechanics it does. 
Certainly, one may revise the above $\psi$-epistemic model so that its predictions may be consistent with those of quantum mechanics for the first order of $1/N$. But it seems difficult to revise the model so that its predictions are consistent with those of quantum mechanics for all orders of $N$. More work needs to be done here. 

Combes et al (2018) also proposed a $\psi$-epistemic model for an A-PM for some observables.  
In the model, the  wave function  is a coherent state of a quantum harmonic oscillator. The Hamiltonian of the system is set to make this state be its nondegenerate ground state. 
Then the system is coupled to a pointer via the usual interaction Hamiltonian $H_I = PA/T$ for a time duration $T$, where $P$ is the conjugate momentum of the pointer variable $X$, and $A$ is a measured quadrature observable.\footnote{Here I use a notation somewhat different from the original one.} 
In the Heisenberg picture, the pointer variable at time $t$ during the A-PM is (up to the first order of 1/T)

\begin{equation}
X(t)=X(0)+{t \over T}\exptt{A}+{1 \over T}[q(0)\sin t+p(0)(1-\cos t)],
\label{m2}
\end{equation}

\noindent where $q(0)$ is the initial position of the system, and $p(0)$ is the initial momentum of the system. 

In this $\psi$-epistemic model for an A-PM, as in the previous $\psi$-epistemic model for a Z-PM, it is still assumed that any observable $A$ of a system has a definite value at any time, which is one of the eigenvalues of $A$, and in particular, the pointer also has a definite position at any time, which is the same as the measured position predicted by quantum mechanics.  
Then, when $T \rightarrow \infty$, we have $X(T)=X(0)+\exptt{A}$, which means that the pointer shift is indeed the result of the A-PM, namely the expectation value of the measured observable.  

However, it can be seen that like the previous $\psi$-epistemic model for a Z-PM, this $\psi$-epistemic model for an A-PM is also inconsistent with quantum mechanics in the predictions about the variance of the measurement result for non-ideal situations in which the measurement interval $T$ is finite.  
According to the model, the variance of the final position of the pointer after the A-PM is

\begin{equation}
Var(x_f)=Var(x_0) +{1 \over T^2}[Var(q_0)\sin^2 T+Var(p_0)(1-\cos T)^2],
\label{}
\end{equation}

\noindent where $Var(q_0)$ is the initial position variance of the system, and $Var(p_0)$ is the initial momentum variance of the system. 
This time the discrepancy is more obvious. 
Quantum mechanics predicts that the variance of the final measured position of the pointer after the A-PM should have the first order term which depends on the initial  measured position of the pointer, while the above model predicts that there is no such a term.

One may also revise the above $\psi$-epistemic model for an A-PM so that its predictions are consistent with those of quantum mechanics for the first order of $1/T$. However, it seems difficult to obtain the consistency, let alone the consistency for all orders of $1/T$. For example, look at the final wave function of the combining system after an A-PM, which is (up to the first order of $1/T$)
\begin{multline}
\ket{t=T} =\ket{\psi} \ket{\phi(x_0+\langle A \rangle)}+ \frac{1}{T} \sum_{m} \frac{1}{E-E_m}\ket{E_m}  \\
\times [\bra{E_m}A\ket{\psi} \ket{\widetilde{\phi}(x_0+\langle A \rangle)}-\E^{\I (E-E_m)T}\bra{\psi}A\ket{E_m} \ket{\widetilde{\phi}(x_0+\langle A \rangle_m)}],\label{}
\end{multline}
where $E$ is the energy of the measured state $\ket{\psi}$, $\ket{E_m}$ are the other energy eigenstates, $E_m$ are the corresponding energy eigenvalues,  $\ket{\widetilde{\phi}(x_0)}$ is a distorted version of the initial pointer wavepacket (see Schlosshauer and Claringbold [2014]), and $\langle A \rangle_m \equiv \bra{E_m}A\ket{E_m}$. 
In order to make the same predictions as quantum mechanics about the variance of the result for the first order of $1/T$, the $\psi$-epistemic model needs to consider the infinitely many energy eigenvalues, $\{E_m\}$, which will appear in the variance of the result predicted by quantum mechanics.

To sum up, I have shown that although the above $\psi$-epistemic models for Z-PMs and A-PMs are consistent with quantum mechanics for ideal situations, namely when $N \rightarrow \infty$ and $T \rightarrow \infty$,\footnote{In the limit $N \rightarrow \infty$ or $T \rightarrow \infty$, there is still an uncertainty in the measured value of $\langle A \rangle$ that comes from the final width of the pointer wavepacket. The width  of the pointer wavepacket after $\delta t$ is $W_{\delta t}=\sqrt{{1 \over 2}(W_0^2+{{\delta t^2}\over{M^2W_0^2}})}$, where $W_0$ is the initial width of the  pointer wavepacket, and $M$ is the mass of the pointer. When the initial width of the pointer wavepacket is small enough and the mass of the pointer is large enough, the final width of the pointer wavepacket after a finite measurement time will be small enough. Thus the uncertainty in the measured value of $\langle A \rangle$ can be made arbitrarily small in principle for ideal PMs.}  they are not fully consistent with quantum mechanics for finite $N$ and $T$. 
In the next section, I will argue that no $\psi$-epistemic models exist for PMs under certain auxiliary assumption about the dynamics of the ontic state. 

\section{A stronger result}

According to the basic result obtained before, 
the wave function of a protected system is determined by the total evolution of the ontic state of the system during a PM. 
In general, the ontic state of the system may be disturbed by the protection procedure and its total evolution may be determined not only by the initial ontic state, but also by the protection procedure. 
Only if we can prove that the wave function of a protected system is determined only by the initial ontic state of the system, not by the protection procedure, can we establish the reality of the wave function. 
The ontic indifference assumption in Hardy's theorem can help us, but it is a too strong assumption. 
In this section, I will suggest a weaker and more natural assumption about the dynamics of the ontic state, and argue that under this auxiliary assumption, the reality of the wave function can be proven in the ontological models framework for PMs. 
 
Consider an ideal PM of an arbitrary observable $A$. 
The initial wave function of the measured system is $\ket{\psi}$. 
As before, the interaction Hamiltonian is given by the usual form $H_I = g(t)PA$, where $g(t)$ is the time-dependent coupling strength of the interaction, which is a smooth function normalized to $\int_0^T g(t)dt =1$ during the measurement interval $T$, and $g(0)=g(T)=0$, and $P$ is the conjugate momentum of the pointer variable $X$. Then the pointer shift after a time $\delta t$ during the PM is: 

\begin{equation}
\Delta x = \exptt{A}\int_0^{\delta t}g(t)dt,
\label{ps}
\end{equation}

\noindent where $\Delta x = \exptt{X}_{\delta t}- \exptt{X}_{0}$, $\exptt{X}_{0}$ is the center of the initial pointer wavepacket, $\exptt{X}_{\delta t}$ is the center of the pointer wavepacket after ${\delta t}$, and $\exptt{A}$ is the expectation value of the measured observable $A$. Here I used the fact that the wave function of the measured system is not changed during the PM. 

When $\delta t = T$ we obtain $\Delta x = \exptt{A}$, namely the result of the PM is the expectation value of the measured observable. 
According to the basic result obtained before, 
the wave function of the protected system is determined by the total evolution of the ontic state of the system during the measurement interval $T$, denoted by $\lambda(t)$, where $t \in [0,T]$. 
Now the key is to notice that when the time-dependent coupling strength $g(t)$ is known, we can also obtain the value of $\exptt{A}$ after any $\delta t > 0$ during the PM, which is $\exptt{A}=\Delta x/\int_0^{\delta t}g(t)dt$.\footnote{Alternatively, we may also adjust the parameters of a PM such as $g(t)$ so that the new measurement time is shorter than the original measurement time. In this case, we may obtain the result $\Delta x=\exptt{A}$ after the PM. However, when the total measurement time is arbitrarily short, it is required that $g(t)$ and the interaction Hamiltonian should be arbitrarily large. My following argument based on the auxiliary assumption cannot apply to this situation.}  
Then, by a similar argument as for the basic result, 
the wave function of the protected system is also determined by the total evolution of the ontic state of the system during the time interval $\delta t$, namely $\lambda(t)$, where $t \in [0,\delta t]$. 

Here is another argument by reduction to absurdity. 
For two protected wave functions $\ket{\psi_1}$ and $\ket{\psi_2}$, choose an observable $A$ whose expectation values in these two states are different. 
If there exists a probability $p>0$ that these two wave functions correspond to the same time evolution of the ontic state of the protected system, $\lambda(t)$, during the time interval $\delta t$, then since $\lambda(t)$ determines the behaviour of the measuring device according to the ontological models framework, 
the pointer shift after $\delta t$ for these two PMs will be the same with a probability not smaller than $p$. 
This leads to a contradiction; the pointer shift after $\delta t$ for these two PMs are 
$\Delta x = \exptt{A}_{\psi_1}\int_0^{\delta t}g(t)dt$ and $\Delta x = \exptt{A}_{\psi_2}\int_0^{\delta t}g(t)dt$, respectively, which are different with certainty since $\exptt{A}_{\psi_1} \neq \exptt{A}_{\psi_2}$ and $g(t)$ is the same for the same measurement of $A$ on the two protected wave functions $\ket{\psi_1}$ and $\ket{\psi_2}$. 

A similar argument can also be given in terms of non-ideal or realistic PMs. For a realistic PM of an observable $A$, there is always a small probability to obtain a result different from $\exptt{A}$, and after any $\delta t > 0$ during the PM there is also a small probability for the pointer shift to be different from the value given by Eq. (\ref{ps}).  
In this case, according to the ontological models framework, the behaviour of the measuring device such as the probabilities for different pointer shifts will be determined by the realistic measuring condition including the ontic state of the measuring device and the measurement interval, as well as by the total evolution of the ontic state of the protected system during the PM. 

Now consider two realistically protected wave functions, and choose an observable whose expectation values in these two states are different. Then we can perform the same measurement of the observable on these two protected states. 
If there exists a probability $p>0$ that these two wave functions correspond to the same evolution of the ontic state of the protected system, $\lambda(t)$, during the time interval $\delta t$, then since the same $\lambda(t)$ yields the same probability distribution of the pointer shift under the same measuring condition according to the ontological models framework, the overlap of the probability distributions of the pointer shift for these two measurements will be not smaller than $p$. 
On the other hand, 
if quantum mechanics is valid, then a realistic condition can always be reached so that the overlap of the probability distributions of the pointer shift after $\delta t$ for these two measurements is smaller than $p$, since when the realistic condition approaches the ideal condition the overlap will approach zero.\footnote{Certainly, this argument will be invalid if quantum mechanics breaks down when reaching certain realistic condition.}  
This leads to a contradiction. 
Therefore, two protected wave functions of a system correspond to different evolution of the ontic state of the system during the time interval $\delta t$. 

Since $\delta t$ can be arbitrarily small in principle, the above result means that 
the wave function of a protected system is determined by the total evolution of the ontic state of the system during an arbitrarily short time interval or an infinitesimal time interval around the initial instant (when the PM starts). 
In order to know whether the wave function of a protected system is determined only by the initial ontic state of the system immediately before the PM, we need a further analysis of the change of the ontic state after the PM starts. 

For a PM such as an A-PM to work, we must turn the protection on completely before the PM can start, and thus the PM and the switching-on of the protection are two different processes occurring during non-overlapping time intervals, namely we have a switching-on of the protection followed by a PM of duration $\delta t$. 
Now, in order to prove the reality of the wave function, I resort to a finiteness assumption, which says that a finite interaction causes a finite rate of change of the ontic state.\footnote{An example in classical mechanics is that a finite force causes a finite acceleration, namely a finite rate of change of velocity. Note that the complete ontic state of a quantum system may be composed of two parts. During a PM, the time evolution of one part is continuous, while the time evolution of the other part is discontinuous. In this case, the finiteness assumption needs to hold true only for the continuous part of the ontic state.} 
Concretely speaking, when a system has a finite interaction with other systems (namely the interaction Hamiltonian is finite), the time evolution of its ontic state is continuous, and the rate of change of the ontic state is finite. 

First, the protection such as the protective potential for an A-PM is a finite interaction between the protected system and the protective setting (during both the switching-on period and the PM), and the PM of an observable is also a finite interaction. 
Next, the switching-on period of the protection and the duration of the PM can be made arbitrarily short.
Then, the finiteness assumption requires that when the switching-on period, denoted by $\tau$, is infinitely short, the difference between the ontic states of the system after the protection and before the protection, $\lambda(\tau)-\lambda(0)$, is also infinitely small, since the rate of change of $\lambda(t)$ is finite during the period according to the assumption. 
Similarly, when the duration of the PM, $\delta t$, is infinitely short, the difference between the ontic states of the system after the PM and before the PM, $\lambda(\tau+\delta t)-\lambda(\tau)$, is also infinitely small. 
Thus, the wave function of the system, which is determined by the total evolution of the ontic state of the system during an infinitesimal time interval after the PM starts, will be determined by the initial ontic state of the system before the PM, $\lambda(0)$. 
In other words, the wave function of an (unprotected) system is determined by the ontic state of the system. 
This proves the reality of the wave function. 

It will be interesting to see whether the reality of the wave function can be proven without resorting to the finiteness assumption, or whether a $\psi$-epistemic model for PMs can be found by rejecting the assumption. I will investigate these issues in future work. 

\section{Conclusion}

Since the discovery of the new method of protective measurement in quantum mechanics by Aharonov, Vaidman and Anandan in 1993, it has been debated whether it implies the reality of the wave function. On the one hand, since protective measurement  can measure the wave function from a single system, it seems tempting and natural to assume that the wave function is a property of the system. On the other hand, since protective measurement must involve a protection procedure related to the wave function of the measured system, it seems also possible that the wave function is not a property of the system, but generated by the evolution of the actual ontic state of the system induced by the protection procedure. 

In this paper, I present a new analysis of the relationship between protective measurements and the reality of the wave function. 
First, I give a simple proof of Hardy's theorem in terms of protective measurements, which shows that when assuming the ontic state of the protected system keeps unchanged during a protective measurement, the wave function must be real, representing the ontic state of a physical system.  
Second, I argue that under a more natural finiteness assumption about the dynamics of the ontic state, protective measurement implies the reality of the wave function in the ontological models framework. 
In addition, I also analyze two suggested $\psi$-epistemic models of a protective measurement. It is shown that although these models can explain the appearance of expectation values of observables in a single measurement, their predictions about the variance of the result of a non-ideal protective measurement are different from those of quantum mechanics.  These analyses may help clarify the contribution of protective measurements to the debate on the reality of the wave function. 

\section*{Acknowledgements}

I would like to thank Lev Vaidman, Matt Pusey, Matt Leifer, Josh Combes, and three anonymous referees for their helpful comments on earlier drafts. This work is supported by the National Social Science Foundation of China (Grant No. 16BZX021). 

\begin{flushright}
\emph{
  Research Center for Philosophy of Science and Technology\\
  Shanxi University\\
  Taiyuan 030006, P. R. China\\
  gaoshan2017@sxu.edu.cn
}
\end{flushright}


\end{document}